PAPER • OPEN ACCESS

# Dihedral beams

To cite this article: Alfonso Jaimes-Nájera 2025 *J. Opt.* **27** 015604

View the article online for updates and enhancements.

# Dihedral beams

Alfonso Jaimes-Nájera

Tecnologico de Monterrey, School of Engineering and Sciences, Ave. Eugenio Garza Sada 2501,
Monterrey, N.L. 64849, Mexico

E-mail: ajaimes@tec.mx

## Abstract

In this work, a group theory-based formulation that introduces new classes of
dihedral-symmetric beams is presented. Our framework leverages the algebraic properties of the
dihedral group of rotations and reflections to transform input beams into closed-form families of
dihedral-invariant wavefields, which will be referred to as dihedral beams. Each transformation
is associated with a specific dihedral group in such a way that each family of dihedral beams
exhibits the symmetries of its corresponding group. Our approach is inspired by one of the
outcomes of this work: elegant Hermite–Gauss beams can be described as a dihedral
interference pattern of elegant traveling waves, a new set of solutions to the paraxial equation
also developed in this paper. Particularly, when taking elegant traveling waves as input beams,
they transform into elegant dihedral beams possessing quasi-crystalline properties and including
features like phase singularities, self-healing, and pseudo-nondiffracting propagation, as well as
containing elegant Hermite and Laguerre–Gauss beams as special cases. Our approach can be
applied to arbitrary scalar and vector input beams and constitutes a general group-theory
formulation that can be extended beyond the dihedral group.

## 1. Introduction

The remarkable properties of structured light have given rise to countless applications that have been exhaustively studied and exploited over the past decades [1–4]. One of the main advantages of structured light is the high number of degrees of freedom that can be controlled. This has led to exciting applications in several areas of physics, ranging from optics, micromanipulation, quantum information, and quantum cryptography, to the detection of gravitational waves [4–9]. For instance, since the discovery that structured light can carry orbital angular momentum [10], the study of their fundamental properties received a new boost, particularly that of Laguerre–Gauss beams [11–18].

Another important example of an optical property of light is self-healing, that is, the ability of a light beam to partially self-reconstruct after encountering an obstacle [19]. Its interest has grown over the last three decades [20, 21]. Its analysis has been done from different perspectives -from considerations of wave optics to ray-wave duality and caustics, among others- and has led to the development of new applications that range from micromanipulation, imaging, to telecommunications [4, 21–26]. Moreover, its concept has even been extended to optical self-reconstruction beyond obstructions and to the preservation of properties such as quantum entanglement [27–29]. However, its impact has also generated new knowledge about structured light. For instance, it has been shown that optical beams that have been widely studied for several decades still have properties that had not been explored until recently [30–42]. Notably, it has been shown that Laguerre– and Hermite–Gauss beams are constituted by the superposition of two and four traveling waves, respectively [33, 37, 38]. Even though those traveling waves are not square-integrable, it was shown that they provide a meaningful physical outlook of Laguerre– and Hermite–Gauss beams, as it presents a quantitative and neat physical description of self-healing [37, 38]. Of course, these are examples of shape-invariant beams with self-healing properties. On the other hand, the self-healing properties of non-structural stable beams, such as elegant Hermite–Gauss beams, are perhaps less evident.

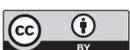

Elegant Hermite–Gauss beams (eHGB), or Siegman's modes, are a variation of the standard Hermite–Gauss beams (sHGB) that result by constructing a more symmetric, or elegant, mathematical configuration of the arguments of the Hermite polynomial and the Gaussian function [43]. This is accomplished by turning the argument into a complex form which completely alters the structure of the sHGB. As a result, its characteristic shape-invariance is lost, and its far-field profile changes into a four-peak far-field pattern [44]. This procedure was also applied to beams with circular geometry, giving rise to elegant Laguerre–Gauss beams (eLGB) [45]. Since then, a large amount of research has been carried out on elegant beams, from studying their properties [44, 46, 47], proposing variants in other geometries [48], introducing partial coherent versions [49], vector and non-paraxial formulations [50], and even studying elegant modes in resonant cavities [18, 51]. This has led to several applications in optical micromanipulation and atmospheric propagation, among others [52–54]. Recently, the non-diffracting and self-healing properties of eHGB and eLGB were studied by comparing them to pseudo-nondiffracting beams [55]. Some important cases of pseudo-nondiffracting beams can be described as interference patterns of plane waves modulated by a Gaussian envelope [56, 57]. This has led to their application in generating optical lattices with kaleidoscopic structures [58–65]. Considering the above, the question of whether the eHGB can also be described by the sum of traveling waves naturally arises. Moreover, if eHGB do indeed result from an interference pattern, could they correspond to a particular case of a more general class of structured interference patterns of light? To the best of our knowledge, no analysis has aimed to answer these questions.

In this work, we show that elegant Hermite–Gauss beams are a specific family of dihedral interference patterns of four elegant traveling waves within a broader class of dihedral beams. By using the complete set of solutions to the Hermite equation, we constructed elegant traveling waves that constitute new solutions to the paraxial Helmholtz equation to the best of our knowledge. We show that these elegant waves behave asymptotically as skew quasi-Gaussian beams (sqGBs), and that can be used to describe, in simple physical terms, optical properties of eHGB, such as self-healing, self-splitting, and pseudo-non-diffracting propagation. To validate our predictions, we have performed numerical experiments on the propagation of the corresponding light beams. Furthermore, we leverage their symmetry properties to establish a formulation, based on the dihedral group of rotations and reflections, that transforms input beams into families of dihedral beams. Particularly, when taking sqGBs as input beams, dihedral interference patterns that contain as special cases the elegant Hermite– and elegant Laguerre–Gauss beams, are formed. In this case, the resulting family of fields will be referred to as elegant dihedral beams. The transverse profiles of each family are invariant under the transformations of the corresponding dihedral group. Consequently, elegant dihedral beams possess a quasi-crystalline transverse profile with specific symmetries, including features like phase singularities, self-healing, and pseudo-nondiffracting propagation properties. Our formulation is general; it is valid for any kind of input beams and is compatible not just with the dihedral group but also with any finite group of linear transformations.

## 2. Elegant traveling-wave solutions to the paraxial equation

The eHGB are solutions to the paraxial Helmholtz equations that can be written in the normalized cartesian coordinates $(x, y, z)$ as [66]

$$eHG_{m,n}(x,y,z) = E_0 \left(\sqrt{\frac{1}{1+iz}}\right)^{m+n+2}$$
$$\times H_m\left(\frac{x}{\sqrt{1+iz}}\right) H_n\left(\frac{y}{\sqrt{1+iz}}\right) \exp\left(-\frac{x^2+y^2}{1+iz}\right), \quad (1)$$

where $n$ and $m$ are integers. The normalized coordinates are defined as $x = x'/w_0$, $y = y'/w_0$, $z = z'/z_L$, where $(x', y', z')$ are the regular cartesian coordinates, $w_0$ is the Gaussian beam waist, $z_L = kw_0^2$ is the diffraction length, $k = 2\pi/\lambda$ is the wavenumber and $\lambda$ the wavelength of the light beam. The function $H_n$ is the Hermite polynomial and is one of the solutions to the Hermite equation that arises when solving the paraxial equation. Their elegance is attributed to the fact that the argument of the Gaussian term is $-1$ times the square of the argument of the Hermite polynomial, namely, $H_m(x)e^{-x^2}$. The algebraic distinction between elegant and standard Hermite–Gauss beams results in significant changes in their optical properties. For instance, unlike sHGB, eHGB are not shape-invariant and produce a four-peaked pattern in the far-field region, as shown in figures 1(a)–(c) [44]. Next, we show that eHGB can be expressed as the superposition of traveling waves and that this formulation provides simple explanations of the above properties on physical grounds.

It has been shown that analyzing the asymptotic behavior of the functions describing the transverse structure of a beam can be helpful in constructing traveling-wave solutions to the paraxial equation [30, 37, 38]. Therefore, we analyze the asymptotics of the product of the Gaussian and the Hermite polynomial that characterizes the eHGB transverse profile, namely,

$$H_n(x)e^{-x^2} \sim A_n \cos(a_n x) e^{-x^2/2}, \quad (2)$$

which applies for large even values of $n$, where $A_n = (-1)^{n/2} n!/(n/2)!$, and $a_n = \sqrt{2n+1}$, with an analogous expression for $n$ odd in terms of the sine function [38, 67]. Notice the 'elegance' of the left-hand side of equation (2); it has the symmetry of the elegant Hermite–Gauss beams. The right-hand side of the latter equation resembles the cosine-Gauss beam, as it has recently been shown [55]. However, rather than focusing solely on this fact, we can interpret from equation (2) that 1D elegant Hermite solutions result from the sum of two traveling-wave solutions; $A_n e^{\pm i a_n x} e^{-x^2/2}$. To construct them, a second linearly independent solution to the Hermite equation behaving asymptotically as $A_n \sin(a_n x) e^{-x^2/2}$ is required. The Neumann-Hermite function $NH_n(x)$ is a second linearly independent solution to the Hermite equation satisfying such requirement, and it has been used to construct Hankel–Hermite functions with which traveling-wave solutions to the paraxial equation can be defined [38]. The construction of elegant traveling-like solutions is now clear;

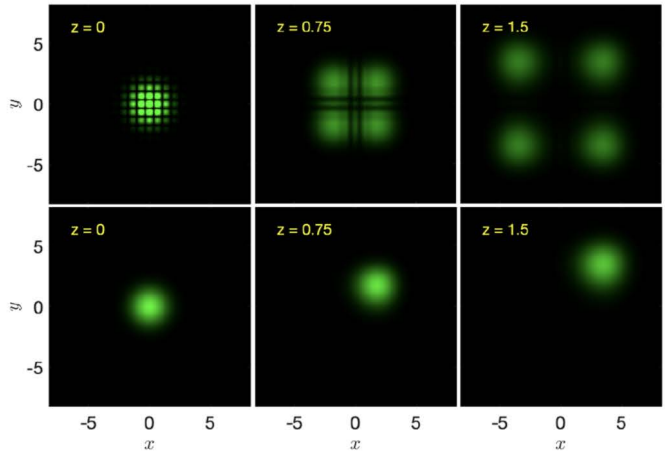

**Figure 1.** Propagation of the elegant Hermite–Gauss beam (first row) and of the elegant traveling wave $eU_{m,n}^{(1,1)}$ (second row) at $z=0$, $z=0.5$, and $z=1$ in normalized coordinates, for $n=m=10$ in both cases.

$$eHH_n^{(1,2)}(x) := [H_n(x) \pm i NH_n(x)] e^{-x^2} \sim c_n e^{\pm i a_n x} e^{-x^2/2}, \quad (3)$$

where the asymptotics hold for any integer $n$ satisfying $n \gg 1$, and * denotes complex conjugation. The expressions of $NH_n$ and $c_n$ are shown in appendix A. The functions $eHH_n^{(1,2)}(x)$ are the elegant version of the Hankel–Hermite functions [38]. Hence, four different elegant traveling-wave solutions to the paraxial equation can be constructed;

$$eU_{m,n}^{(j,k)}(x,y,z) = E_0 \left(\sqrt{\frac{1}{1+iz}}\right)^{m+n+2}$$
$$\times eHH_m^{(j)}\left(\frac{x}{\sqrt{1+iz}}\right) eHH_n^{(k)}\left(\frac{y}{\sqrt{1+iz}}\right), \quad (4)$$

where $j,k = 1,2$. Their asymptotic behavior for $m,n \gg 1$ can be expressed, according to equation (3), as

$$eU_{m,n}^{(j,k)}(x,y,z) \sim A_{m,n}(z) e^{-i[(-1)^j a_n x_z + (-1)^k a_m y_z]} e^{-(x_z^2 + y_z^2)/2}, \quad (5)$$

where $A_{m,n}(z) = E_0 c_m c_n (1+iz)^{-(m+n+2)/2}$, and $x_z = x/\sqrt{1+iz}$, $y_z = y/\sqrt{1+iz}$. The traveling-wave nature of these beams is now evident. Remarkably, in analogy with sHGB [38], the eHGB are the interference pattern of the above four elegant traveling waves;

$$eU_{m,n}^{(1,1)} + eU_{m,n}^{(1,2)} + eU_{m,n}^{(2,1)} + eU_{m,n}^{(2,2)} = 4 eHG_{m,n}, \quad (6)$$

which answers the first question posed in the introduction of this work.

Elegant traveling waves $eU_{m,n}^{(j,k)}$ are square-integrable. This contrasts with the non-square-integrability of the traveling waves of the sHGB [38]. Interestingly, as shown in appendix A, their square-integrability is due to the symmetry between the arguments of the Gaussian and the Hermite polynomial, or in other words, to their 'elegant' structure. We will now analyze their physical properties to exploit their symmetry properties later.

### 2.1. Physical analysis of elegant traveling waves

The asymptotic expansions previously developed contain physical information about the elegant traveling waves. According to equation (5), it is clear that at least near the $z=0$ transverse plane, the elegant wave $eU_{m,n}^{(1,1)}$ propagates asymptotically as a Gaussian beam in the direction of the first quadrant of the transverse plane. This behavior is maintained for $z > 0$, as shown in figure 1. Therefore, the elegant traveling wave propagates towards the first quadrant of the transverse plane as a sqGB. Initially, it has a beam width $w(0) = \sqrt{2}$. Later, in the far-field region, the elegant traveling wave $eU_{m,n}^{(1,1)}$ continues to behave asymptotically as a sqGB. Indeed, as it is proved in appendix B, for large values of $z$, the asymptotics of the elegant traveling waves can be expressed as

$$eU_{m,n}^{(\pm,\pm)}(\mathbf{r}) \sim E_{m,n}^{(\pm)} \frac{e^{i(x^2+y^2)/z}}{1+iz} \left(\frac{1-iz}{1+iz}\right)^{\frac{n+m}{2}}$$
$$\times \exp\left(\frac{-\left(\pm x - z\sqrt{n/2}\right)^2 - \left(\pm y - z\sqrt{m/2}\right)^2}{\left(z/\sqrt{2}\right)^2}\right), \quad (7)$$

where $E_{m,n}^{(\pm)}$ is an amplitude constant, and the superscript $+$ $(-)$ in equation (7), corresponds to the superscript 1 (2) in equation (4). Then, it can be concluded that the elegant traveling wave far-field pattern has, at first order, a single Gaussian peak, with a beam width $w(z) = z/\sqrt{2}$, propagating along the line given by $x = \pm\sqrt{n/2}z$, $y = \pm\sqrt{m/2}z$. In summary, each elegant wave propagates as a sqGB towards one of the quadrants of the transverse plane along the propagation axis (see figures 1(d)–(f)).

## 3. Rotations, reflections, and the dihedral group

The Hermite polynomials $H_n$ are well-known for having definite parity. It is easy to prove that the Neumann-Hermite function $NH_n$ also has a definite parity. Specifically, for even (odd) $n$, $H_n$ is an even (odd) function, while $NH_n$ is an odd (even) function. This implies that the elegant Hankel–Hermite functions can be transformed into each other through a reflection of the $x$-axis;

$$eHH_n^{(1)}(-x) = (-1)^n eHH_n^{(2)}(x). \qquad (8)$$

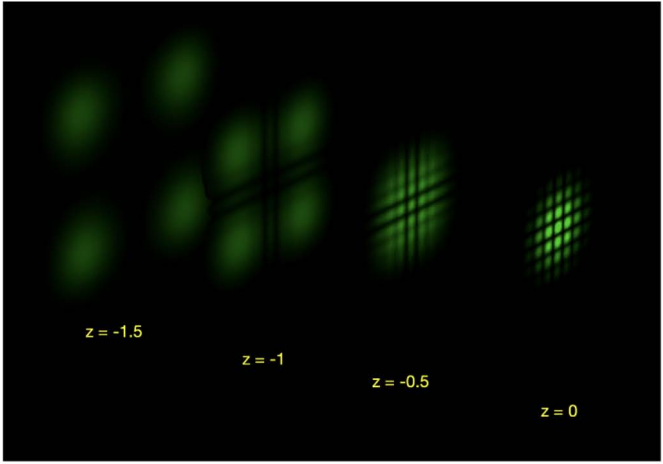

**Figure 2.** 3D perspective of the propagation of the elegant Hermite–Gauss beam with $n = m = 10$, in normalized coordinates.

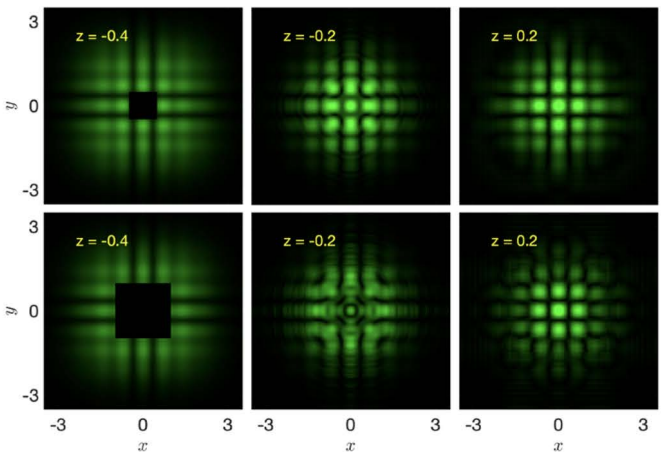

**Figure 3.** Propagation of an obstructed elegant Hermite–Gauss beam at $z = -0.4$, in normalized coordinates. Each row shows a different-sized obstruction.

An interesting consequence is that a single elegant traveling wave can be transformed into the remaining three by applying appropriate reflections and rotations over one of them, for instance, over $eU_{m,n}^{(1,1)}$. This reduces the analysis of the traveling waves of an eHGB to only one. Indeed, according to equations (4) and (8), $eU_{m,n}^{(2,1)}$ is obtained after a reflection of the $x$-axis; $eU_{m,n}^{(2,1)}(x,y,z) = (-1)^n eU_{m,n}^{(1,1)}(-x,y,z)$, whereas $eU_{m,n}^{(2,2)}$ is obtained after a rotation of $\pi$ radians; $eU_{m,n}^{(2,2)}(x,y,z) = (-1)^{n+m} eU_{m,n}^{(1,1)}(-x,-y,z)$, while $eU_{m,n}^{(1,2)}$ is obtained after a rotation of $\pi$ radians followed by a reflection of the $x$-axis; $eU_{m,n}^{(1,2)}(x,y,z) = (-1)^m eU_{m,n}^{(1,1)}(x,-y,z)$. The above is evident in figure 1(c). Of course, $eU_{m,n}^{(1,1)}$ is obtained after applying twice the rotation of $\pi$ radians, i.e. after applying the identity operation on $eU_{m,n}^{(1,1)}$. These four operations are precisely the elements of the dihedral group of rotations and reflections $D_2 = \{R, R^2, S, SR\}$, or symmetry group of a 2-polygon, where $R$ and $S$ are, respectively, the rotation of $\pi$ radians about the $z$-axis and the reflection of the $x$-axis, whereas $SR$ is the composition of $S$ and $R$ [68].

Given that the rotations and reflections act on the coordinates, let us denote the action of the corresponding transformation on a function $f(\mathbf{r})$, as

$$\mathbf{R}f(\mathbf{r}) = f(R\mathbf{r}), \qquad \mathbf{S}f(\mathbf{r}) = f(S\mathbf{r}). \qquad (9)$$

Therefore, according to equation (6) and the above discussion, the eHGB can be written as

$$\left[(-1)^n \mathbf{S} + (-1)^{n+m} \mathbf{R} + (-1)^m \mathbf{SR} + \mathbf{R}^2\right] \times eU_{m,n}^{(1,1)} = 4eHG_{m,n}. \qquad (10)$$

This sets a natural platform to extend the family of elegant beams by using the structure of the general dihedral group $D_\nu$, $\nu$ being a natural number. This group is conformed by $\nu$ rotations and $\nu$ reflections under which a $\nu$-polygon is symmetrical. In other words, $D_\nu = \{R_1, R_2, \cdots, R_\nu, S_1, S_2, \cdots, S_\nu\}$, where its elements can be described in matrix form as,

This eHGB formulation allows a straightforward physical description of their propagation characteristics; four counter-propagating sqGB propagate from the far-field $z < 0$ region and converge near the optical axis, giving rise to the familiar near-field eHGB profile as an interference pattern (figure 2). This pattern remains unchanged over a finite propagation distance as long as the four sqGB interfere, giving rise to pseudo-nondiffracting properties. After some propagation distance, the four Gaussian peak pattern re-emerges, that is, the eHGB self-splits. Furthermore, if an eHGB is partially obstructed at a transverse plane in which its four sqGB are still converging, its familiar near-field profile will be self-reconstructed by counterpropagation of the sqGB. Indeed, figure 3 shows a partially blocked eHGB at a transverse plane $z < 0$, whose near-field pattern is partially reconstructed after a propagation distance at which the four counterpropagating sqGB converge. This demonstrates that, although eHGBs are not structurally stable beams upon propagation, they have self-healing properties due to their structured counterpropagating nature.

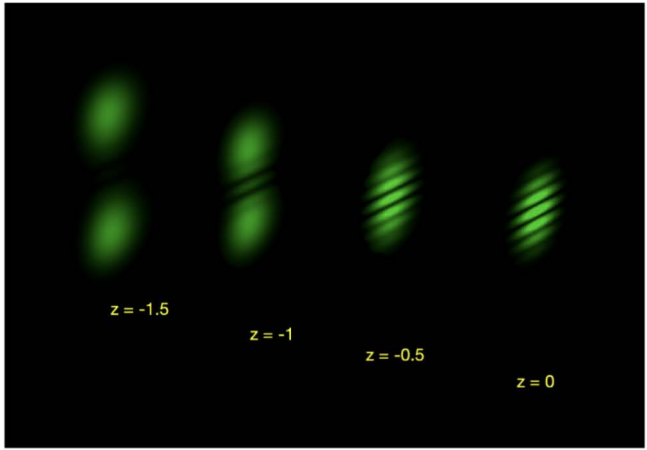

(a)

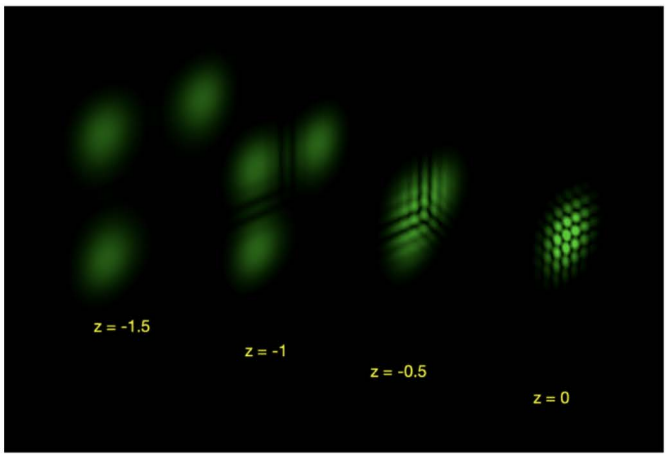

(b)

**Figure 4.** 3D visualization of the propagation of (a) $\left(\mathbf{R}^2 + \mathbf{SR}\right) \mathrm{e}U_{m,n}^{(1,1)}$, and (b) $\left(\mathbf{R}^2 + \mathbf{R} + \mathbf{SR}\right) \mathrm{e}U_{m,n}^{(1,1)}$ (in the notation of equation (10)), for $m = n = 10$, in both cases. Note that the Gaussian peak of $\mathrm{e}U_{m,n}^{(1,1)}$ lies in the third quadrant of the transverse plane for $z < 0$, and that $\mathbf{R}^2$ is the identity operation.

$$R_j = \begin{pmatrix} \cos\frac{2\pi j}{\nu} & -\sin\frac{2\pi j}{\nu} & 0 \\ \sin\frac{2\pi j}{\nu} & \cos\frac{2\pi j}{\nu} & 0 \\ 0 & 0 & 1 \end{pmatrix},$$
$$S_j = \begin{pmatrix} \cos\frac{2\pi j}{\nu} & \sin\frac{2\pi j}{\nu} & 0 \\ \sin\frac{2\pi j}{\nu} & -\cos\frac{2\pi j}{\nu} & 0 \\ 0 & 0 & 1 \end{pmatrix}. \quad (11)$$

On the other hand, the subset of rotations, $d_\nu = \{R_1, R_2, \cdots, R_\nu\}$, is a subgroup of the dihedral group. In fact, any subgroup of $D_\nu$ is either a group of rotations or another dihedral group [68].

With these operations, a diverse kind of elegant interference patterns can be produced. Some examples are shown in figures 4 and 5. However, dihedral patterns with specific symmetries and kaleidoscopical properties can be introduced by leveraging the algebraic properties of the dihedral group.

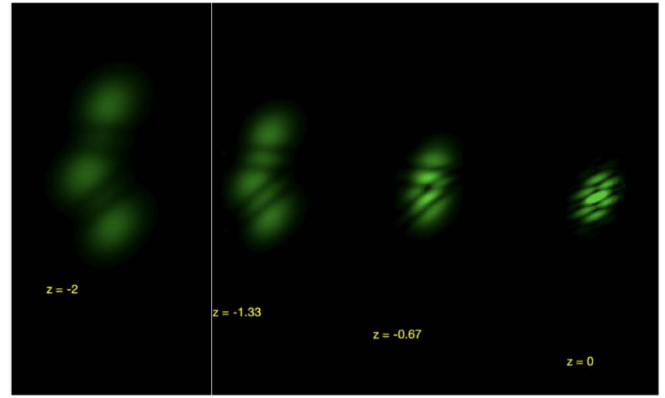

(a)

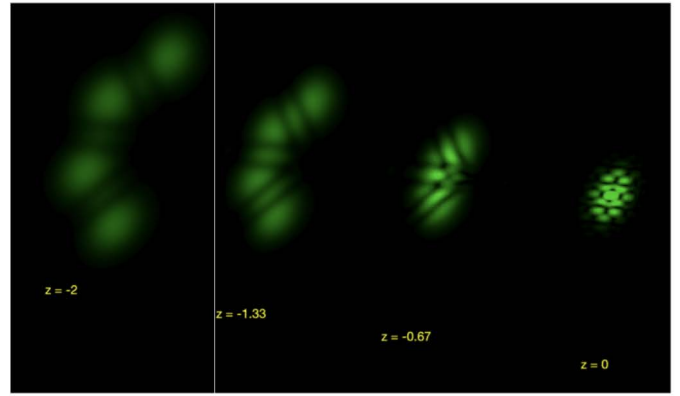

(b)

**Figure 5.** 3D visualization of the propagation of (a) $\left(\mathbf{R}_1 + \mathbf{R}_2 + \mathbf{R}_8\right) \mathrm{e}U_{m,n}^{(1,1)}$, and (b) $\left(\mathbf{R}_1 + \mathbf{R}_2 + \mathbf{R}_3 + \mathbf{R}_8\right) \mathrm{e}U_{m,n}^{(1,1)}$ (in the notation of equation (11)), for $m = n = 10$, in both cases. Note that the Gaussian peak of $\mathrm{e}U_{m,n}^{(1,1)}$ lies in the third quadrant of the transverse plane for $z < 0$.

## 4. Dihedral beam formulation

Consider a finite group of linear transformations, $\{\mathbf{T}_j\}_{j=1,\cdots,\nu}$, that act on a given function space and whose group operation is the composition of transformations. Then, as it is proved in appendix C, **the sum of all transformations of a group acting on a given function $f(\mathbf{r})$, denoted as $\sum_{j=1}^{\nu} \mathbf{T}_j f(\mathbf{r})$, is invariant under any transformation of the group.** This mathematical statement establishes the framework of this paper.

We are particularly interested in the transformations of the dihedral group and its subgroups when taking $\mathrm{e}U_{m,n}^{(1,1)}$ as input function. Let us define the following families of elegant beams

$$\mathrm{e}Dih_{m,n,\nu}^{R+S} = \frac{1}{2\nu} \sum_{j=1}^{\nu} (\mathbf{R}_j + \mathbf{S}_j) \mathrm{e}U_{m,n}^{(1,1)}, \quad (12)$$

$$\mathrm{e}Dih_{m,n,\nu}^{R} = \frac{1}{\nu} \sum_{j=1}^{\nu} \mathbf{R}_j \mathrm{e}U_{m,n}^{(1,1)}. \quad (13)$$

Therefore, $eDih_{m,n,\nu}^{R+S}$ is invariant under any rotation and any reflection in $D_\nu$, while $eDih_{m,n,\nu}^{R}$ is invariant under any rotation in $d_\nu$. Therefore, each dihedral group $D_\nu$ is mapped into the family $eDih_{m,n,\nu}^{R+S}$, which is invariant under $D_\nu$. An analogous mapping is also established between the groups $d_\nu$ and the families $eDih_{m,n,\nu}^{R}$. Since all the subgroups of the dihedral group are either dihedral or rotation groups [68], this mapping encompasses all possible algebraic structures of the dihedral groups.

The families of beams in equations (12) and (13) are dihedral interference patterns of sqGBs. They exhibit rich transverse intensity and phase distributions, as shown in figure 6. We will refer to them as elegant dihedral beams of the first (eDB1) and second kind (eDB2), respectively. Figure 7 shows a 3D visualization of the propagation of the eDB1 for $\nu = 8$, in which it is evident the propagation of eight sqGB on a circle in the transverse plane that converge to form a kaleidoscopic pattern. Since the elegant dihedral beams are written in closed analytical form for any propagation distance, their optical properties can be examined in detail. For instance, elegant dihedral beams possess pseudo-nondiffracting, self-splitting, and self-healing properties simply because they are composed of counterpropagating sqGB. This is shown in figures 8 and 9. Our approach is also useful to conclude that elegant dihedral beams' phase structure includes phase singularities, as is evident in figure 10. This shows that elegant dihedral beams can be considered vortex beams as well.

Figure 6 shows some cases that resemble eHGB. Indeed, it can be proved that when $eU_{m,n}^{(1,1)}$ is selected as input beam, $eDih_{m,n,2}^{R+S} = eDih_{m,n,4}^{R+S} = eDih_{m,n,4}^{R} = eHG_{m,n}$, for $n,m$ even integers. On the other hand, when at least one of the parameters $m,n$ is odd, the corresponding eHGB does not belong to any of the families represented in equations (12) and (13), even though their transverse intensity profile is symmetrical under the transformations of the dihedral group $D_2$. This is because their phases are not invariant. Indeed, $\mathbf{R}eHG_{m,n} = (-1)^{m+n}eHG_{m,n}$, $\mathbf{S}eHG_{m,n} = (-1)^{m}eHG_{m,n}$ and $\mathbf{SR}eHG_{m,n} = (-1)^{n}eHG_{m,n}$, where $\mathbf{R}$ and $\mathbf{S}$ are the elements of $D_2$. This clearly shows that the resulting symmetry of the elegant dihedral beams encompasses both their amplitude and phase structures. Nevertheless, in the next section, we show that eHGB belongs to a modified version of elegant dihedral beams.

The diversity of dihedral patterns our framework produces extends beyond those shown above. When the input beam is replaced by a rotation of $eU_{m,n}^{(1,1)}$, the transverse profiles of the resulting elegant dihedral beam family change while their symmetry is conserved. Taking a rotation of $eU_{m,n}^{(1,1)}$ as input of an eDB2 (equation (13)) has the effect of simply rotating the original eDB2. This is because the group of rotations about the z-axis is abelian; that is, rotations are commutative, which makes it irrelevant whether the rotation of the input beam is operated before or after the transformations of the dihedral group in equation (13). The same happens if we take a reflection of $eU_{m,n}^{(1,1)}$ as input, since $R_j S_k = (S_k)^{-1} R_j$, for any $j,k=1,\cdots,\nu$. However, when considering an initial rotation of $eU_{m,n}^{(1,1)}$ as input of an eDB1, the variety of dihedral patterns increases. This is because the dihedral group $D_\nu$ is non-abelian for $\nu > 2$; i.e. their elements do not commute in general. Figure 11 shows dihedral beam patterns when taking as input beams different rotations of $eU_{m,n}^{(1,1)}$. Thus, by rotating the input beam through an arbitrary angle between 0 and $2\pi$, a continuous diversity of quasi-crystalline patterns with dihedral symmetry can be generated, increasing the variety of kaleidoscopic light structures even more, and enabling the dihedral beam profile to be controlled through the rotation angle of the input beam. This shows how the algebraic properties of the dihedral group can be leveraged to define and control a large variety of dihedral interference patterns, which can be important for practical purposes [64].

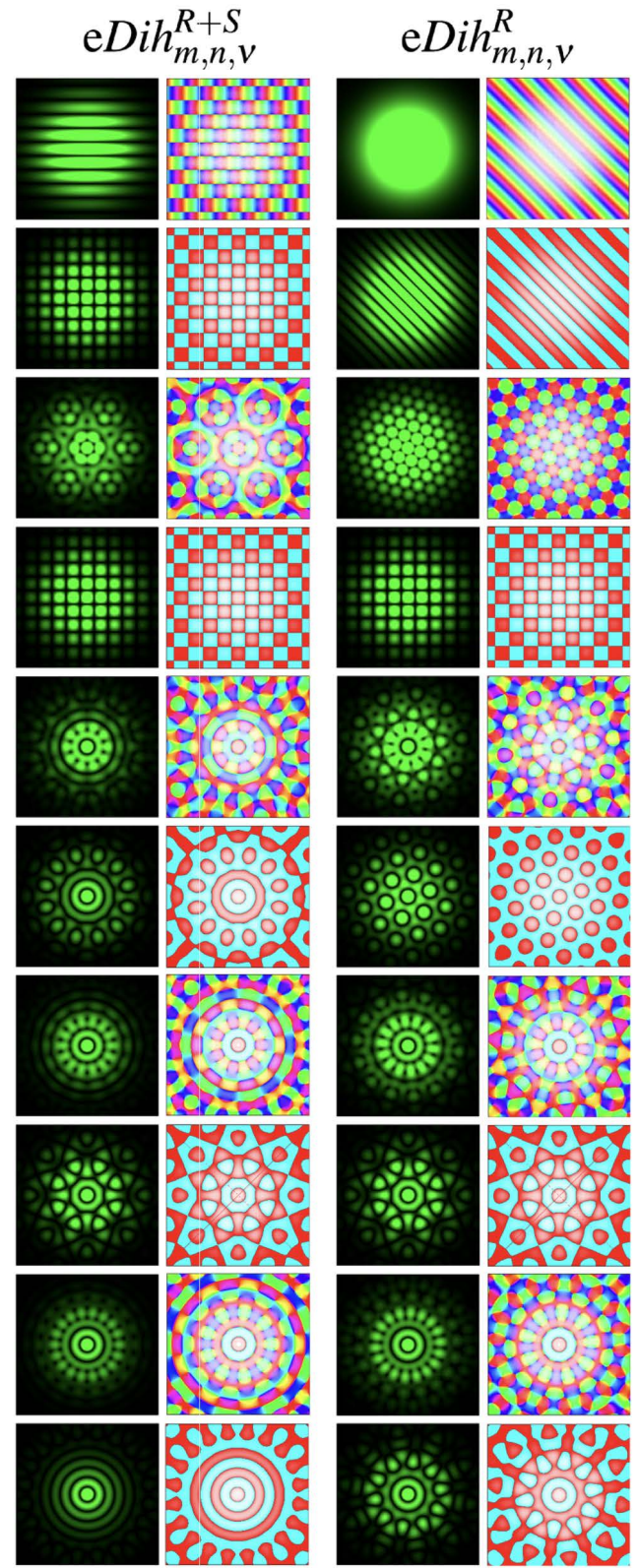

**Figure 6.** Transverse amplitude and phase distributions of the eDB1, $eDih_{m,n,\nu}^{R+S}$ (first two columns, resp.), and eDB2, $eDih_{m,n,\nu}^{R}$ (last two columns, resp.), for $m = n = 20$, at $z = 0$. The first row corresponds to $\nu = 1$, and so on, until the tenth row corresponds to $\nu = 10$, where $\nu$ determines the dihedral group $D_\nu$. The plot range of all the figures is $(-2.5, 2.5)$, in normalized units, for both axes.

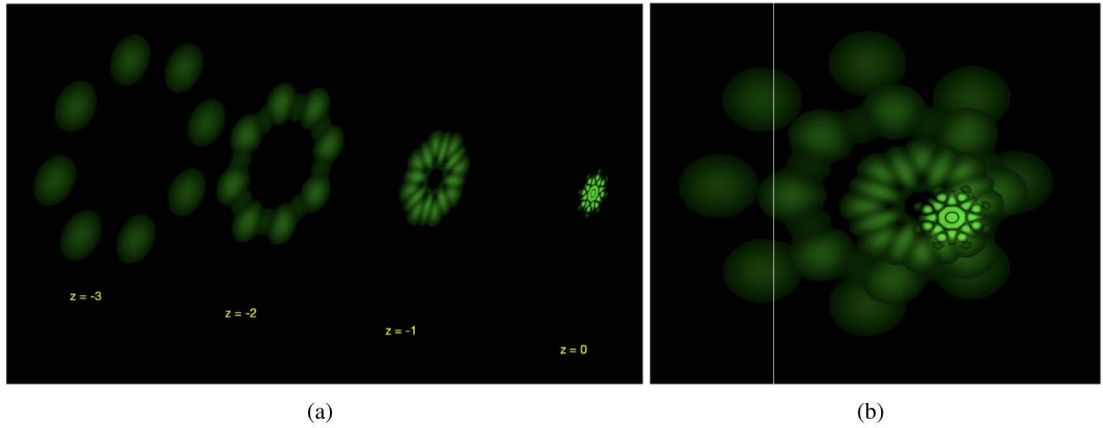

**Figure 7.** 3D visualizations of the propagation of the eDB1, $eDih_{m,n,\nu}^{R+S}$, for $m=n=10$, and $\nu=8$.

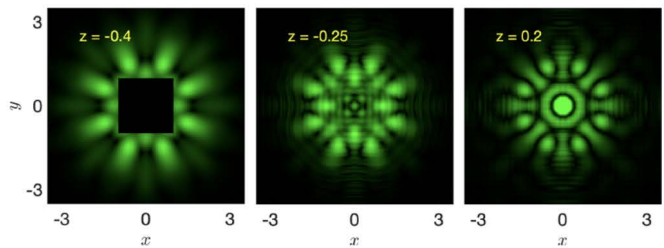

**Figure 8.** Self-healing process of the eDB1, $eDih_{m,n,\nu}^{R+S}$, for $\nu=8$ and $m=n=10$, when obstructed at $z=-0.4$.

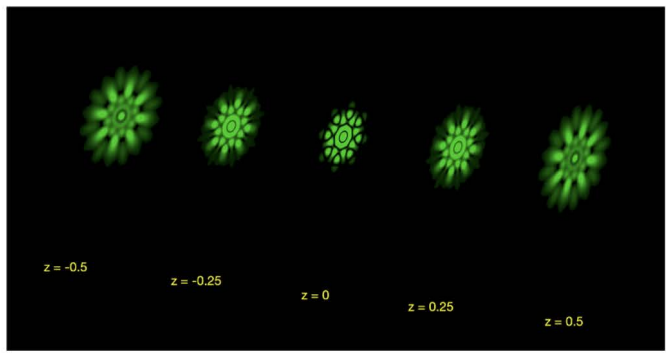

**Figure 9.** 3D perspective of the propagation of eDB1, $eDih_{m,n,\nu}^{R+S}$, for $m=n=10$, and $\nu=8$. Pseudo-nondiffracting propagation properties can be observed over a diffraction length in normalized units.

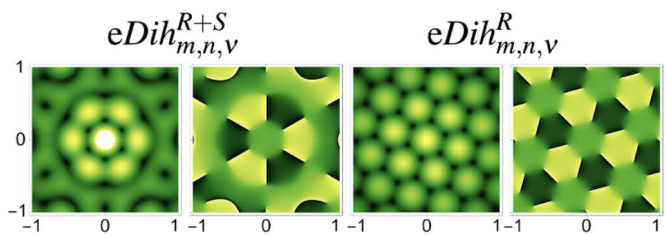

**Figure 10.** Detail of the amplitude and phase distributions of the eDB1 and eDB2, for $\nu=3$, and $m=n=20$, at $z=0$.

The dihedral patterns shown in figure 6 resemble kaleidoscopic structures found in optical lattices [58, 61, 69]. This is because, among the various methods of generating optical lattices, one of particular importance is based on plane wave interference [58–65, 69], resembling the interference of sqGBs. However, to the best of our knowledge, our approach is the first to leverage the properties of the dihedral group to systematically define and classify a large variety of quasi-crystalline light patterns based on the dihedral symmetries of their transverse intensity and phase distributions. Furthermore, this approach provides naturally closed-form expressions of elegant dihedral beams, enabling the analytical study of their propagation properties, a recent area of exploration for kaleidoscopic optical patterns [69].

Although there are various experimental methods for implementing this kind of beams [59, 64, 69], they can also be generated using computer-generated holograms displayed on spatial light modulators [70–74], as this technique has been demonstrated in the generation of very high-order optical vortex beams [74].

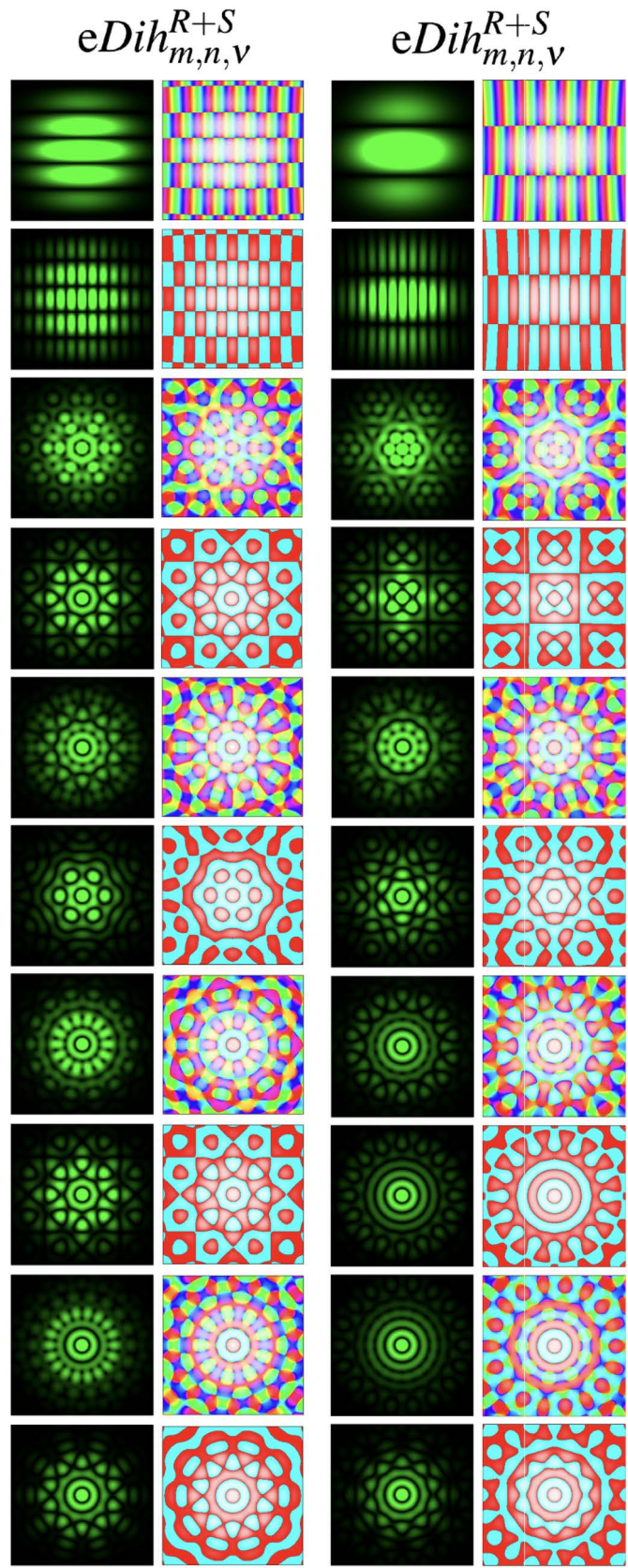

**Figure 11.** Transverse amplitude and phase distributions of the eDB1, $eDih_{m,n,\nu}^{R+S}$, where $U_{m,n}^{(1,1)}$ has been initially rotated by $\pi/8$ (first two columns, resp.), and by $3\pi/16$ radians (last two columns, resp.) around the $z$-axis, for $m = n = 20$, at $z = 0$. As elegant dihedral beams, their dihedral symmetry is preserved. The first row corresponds to $\nu = 1$, and so on, until the tenth row corresponds to $\nu = 10$, where $\nu$ determines the dihedral group $D_\nu$. The plot range of all the figures is $(-2.5, 2.5)$, in normalized units, for both axes.

As the superposition of sqGB conforms elegant dihedral beams, their far-field intensity pattern comprises quasi-Gaussian peaks on a circle (see figure 7). This resembles a discrete, Gaussian-modulated version of the integral used by Durnin to define non-diffracting beams, namely, $W(\mathbf{r};\mathbf{k}) = \exp(ik_z z)\int_0^{2\pi} A(\varphi)\exp[ik_t(x\cos\varphi + y\sin\varphi)]d\varphi$ [57, 69, 75, 76]. In fact, the expression of eDB2, equation (13), corresponds to a Riemann sum whose limit is $W$ for $A(\varphi) = 1$, as $\nu \to \infty$. As we will see next, this perspective is useful for obtaining analytical expressions of the limiting cases $\nu \to \infty$ and for introducing modified versions of the elegant dihedral beams that increase the number of kaleidoscopic interference patterns.

## 5. Modified dihedral beams

More families of dihedral beams can be defined by introducing modifications to the original ones. Let us define the modified elegant dihedral beams of the first (mDB1) and second kind (mDB2) as

$$\mathrm{m}Dih^{R+S}_{m,n,\nu} = \frac{1}{2\nu}\sum_{j=1}^{\nu}(a_j\mathbf{R}_j + b_j\mathbf{S}_j)\mathrm{e}U^{(1,1)}_{m,n}, \quad (14)$$

$$\mathrm{m}Dih^{R}_{m,n,\nu} = \frac{1}{\nu}\sum_{j=1}^{\nu}c_j\mathbf{R}_j\mathrm{e}U^{(1,1)}_{m,n}, \quad (15)$$

respectively, where $a_j$, $b_j$ and $c_j$ are constants. In general, the modified elegant dihedral beams are not invariant under the transformations of the dihedral group. According to equation (10), for any pair of integers $m,n$, the family of eHGB is a specific family of mDB1 with $\nu = 2$. In other words, the whole family of eHGB is a subfamily of modified elegant dihedral beams.

On the other hand, mDB2 can be considered a discrete, 'elegant' version of the integral $W$. In this sense, let us consider the families of modified dihedral beams with $a_j = b_j = c_j = e^{il\varphi_j} = e^{il(2\pi j/\nu)}$, namely, $\mathrm{m}Dih^{R}_{m,n,\nu,l} = \frac{1}{\nu}\sum_{j=1}^{\nu}e^{il(2\pi j/\nu)}\mathbf{R}_j\mathrm{e}U^{(1,1)}_{m,n}$ and $\mathrm{m}Dih^{R+S}_{m,n,\nu,l} = \frac{1}{2\nu}\sum_{j=1}^{\nu}e^{il2\pi j/\nu}(\mathbf{R}_j + \mathbf{S}_j)\mathrm{e}U^{(1,1)}_{m,n}$. Figure 12 shows their transverse amplitude and phase structures. Interestingly, the transverse intensity profile of $\mathrm{m}Dih^{R}_{m,n,\nu,l}$ is invariant under rotations $\mathbf{R}_j \in d_\nu$, since $d_\nu$ is an abelian group. However, their phase structure is not invariant since the rotation $\mathbf{R}_j$ induces a global phase shift. In contrast, the symmetry of the family $\mathrm{m}Dih^{R+S}_{m,n,\nu,l}$ is broken; it is not invariant under rotations since the dihedral group $D_\nu$ is non-abelian for $\nu > 2$. In any case, the modified elegant dihedral beams extend the variety of kaleidoscopic patterns of the original elegant dihedral beams, as shown in figures 12 and 13.

The case of large values of $\nu$ can be studied from the perspective of the integral $W$; the expression of $\mathrm{m}Dih^{R}_{m,n,\nu,l}$ is a Riemann sum that converges to the integral $W$ in the limit $\nu \to \infty$. Therefore, according to the approximation in equation (5), one can obtain that $\lim_{\nu\to\infty}\mathrm{m}Dih^{R}_{m,n,\nu,\mu}(x,y,z) \sim B_{m,n,l}(1+iz)^{-\frac{n+m+2}{2}}J_l(2\sqrt{Nr^2/(1+iz)})\exp(-r^2/[2(1+iz)] - il\phi)$, where $a_n = \sqrt{2n+1}$, $c(n)$ is defined in appendix B, $B_{m,n,l}(z) = (i)^{|l|}(a_n + ia_m)c_n c_m/\sqrt{a_n^2 + a_m^2}$, $N = (n+m+1)/2$, and $r^2 = x^2 + y^2$, $\phi = \arctan y/x$. Analogous results have been found relating Bessel beams to quasi-periodic structures of interference patterns [64]. On the other hand, as Bessel functions behave asymptotically as Laguerre polynomials [31, 37], after some algebra, we can finally write

$$\lim_{\nu \to \infty} \mathrm{m}Dih^R_{m,n,\nu,l}(x,y,z) \sim A_{m,n,l}(z) r^{|l|} L^{|l|}_{\frac{m+n-|l|}{2}}\left(\frac{r^2}{1+iz}\right)$$
$$\times e^{-r^2/(1+iz)} e^{-il\phi}, \quad (16)$$

where $A_{m,n,l}(z) = B_{m,n,l} N^{|l|/2}(1+iz)^{-\frac{m+n+|l|+2}{2}} \Gamma[1+\frac{1}{2}(m+n-|l|)]/\Gamma[1+\frac{1}{2}(m+n+|l|)]$. Remarkably, this approximation applies not only in the near-field, but for any propagation distance $z$. Therefore, the eLGB with radial number $\frac{1}{2}(m+n-|l|)$ and topological charge $l$ is the limiting case of the family $\mathrm{m}Dih^R_{m,n,\nu,l}$. Unlike their standard counterparts, eLGB admit semi-integer and even real values of the radial number while preserving their square integrability. Particularly, for even values of $m,n$, the modified dihedral beams $\mathrm{m}Dih^R_{m,n,\nu,l=0}$ transit from eHGB (at $\nu = 2$) to eLGB with integer radial number $\frac{1}{2}(m+n)$ and $l = 0$ as $\nu \to \infty$. Analogously, it can be proved that

$$\lim_{\nu \to \infty} \mathrm{m}Dih^{R+S}_{m,n,\nu,l}(x,y,z) \sim A_{m,n,l}(z) r^{|l|} L^{|l|}_{\frac{m+n-|l|}{2}}\left(\frac{r^2}{1+iz}\right)$$
$$\times e^{-r^2/(1+iz)} \cos(l\phi). \quad (17)$$

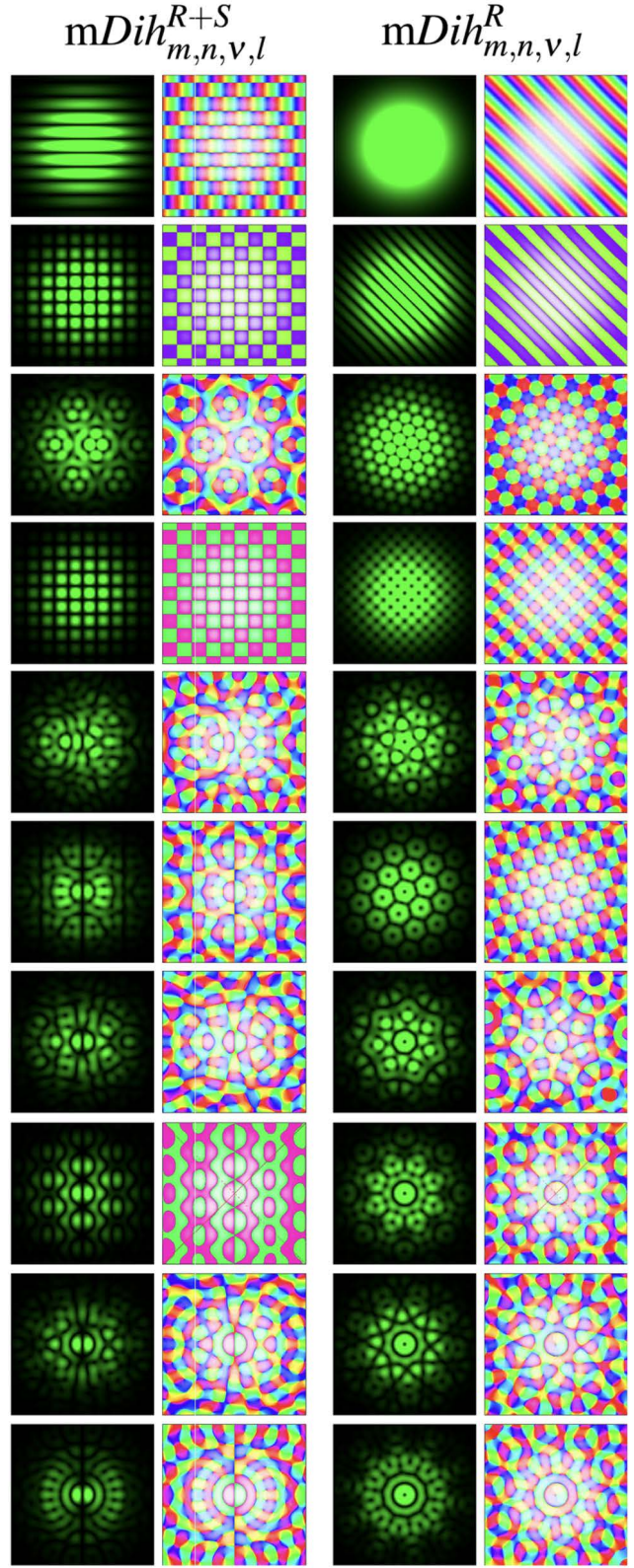

**Figure 12.** Transverse amplitude and phase distributions of the mDB1, $\mathrm{m}Dih^{R+S}_{m,n,\nu,l}$ (first two columns, resp.), and of mDB2, $\mathrm{m}Dih^{R}_{m,n,\nu,l}$ (last two columns, resp.), for $l=1$, $m=n=20$, at $z=0$. Some interference patterns of mDB1 exhibit broken symmetry. The first row corresponds to $\nu=1$, and so on, until the tenth row corresponds to $\nu=10$, where $\nu$ determines the dihedral group $D_\nu$. The plot range of all the figures is $(-2.5, 2.5)$, in normalized units, for both axes.

Figure 14 shows the transition of the elegant dihedral beams towards the limiting eLGB as $\nu$ increases. To the best of our knowledge, the relationship between elegant beams and kaleidoscopical optical structures has not been discussed previously in the literature.

Various families of modified dihedral beams can be created by choosing appropriate constants $a_j$, $b_j$, and $c_j$. For instance, inspired by the integral $W$, modified dihedral beams transiting to Helmholtz–Gauss beams, such as Mathieu–Gauss or Parabolic–Gauss beams, can be constructed, among others [57, 59].

## 6. Concluding remarks

In this paper, we have developed a general group theory-based formulation that leverages the algebraic properties of the dihedral group of rotations and reflections to transform input beams into families of dihedral-symmetric beams. Our framework maps each dihedral group into a family of dihedral beams possessing the symmetries of the corresponding group. Elegant dihedral beams were analytically derived as a special case in which elegant traveling waves are taken as input beams. The resulting wavefields exhibit vortex quasi-crystalline intensity and phase patterns with specific dihedral symmetries. Our approach allows the generation and control of a continuous range of different dihedral beams, all of which are invariant under the transformations of the associated dihedral group. Our formulation provides a systematic method for producing a wide variety of quasi-crystalline and kaleidoscopic light patterns classified according to their symmetries. Moreover, by relaxing symmetry restrictions, modified versions of dihedral beams were introduced, increasing even more the variety of dihedral optical structures that can be constructed. To the best of our knowledge, neither dihedral beams nor elegant traveling waves have been presented in the literature before. We showed that the latter behave asymptotically as skew quasi-Gaussian beams and that they provide a clear, physical basis to understand and describe propagation characteristics of elegant beams such as self-healing, self-splitting, and pseudo-nondiffracting propagation, properties of which elegant dihedral beams inherit. Our approach is inspired by one of the findings of this work: elegant Hermite–Gauss beams are a dihedral interference pattern of elegant traveling waves. In fact, we showed that elegant Hermite–Gauss beams are a specific family of modified elegant dihedral beams. Furthermore, a family of transition beams between even-parameter elegant Hermite-Gauss and orbital angular momentum-carrying elegant Laguerre–Gauss beams was constructed. This shows a relationship between elegant beams and kaleidoscopic patterns of light that has gone unnoticed until now.

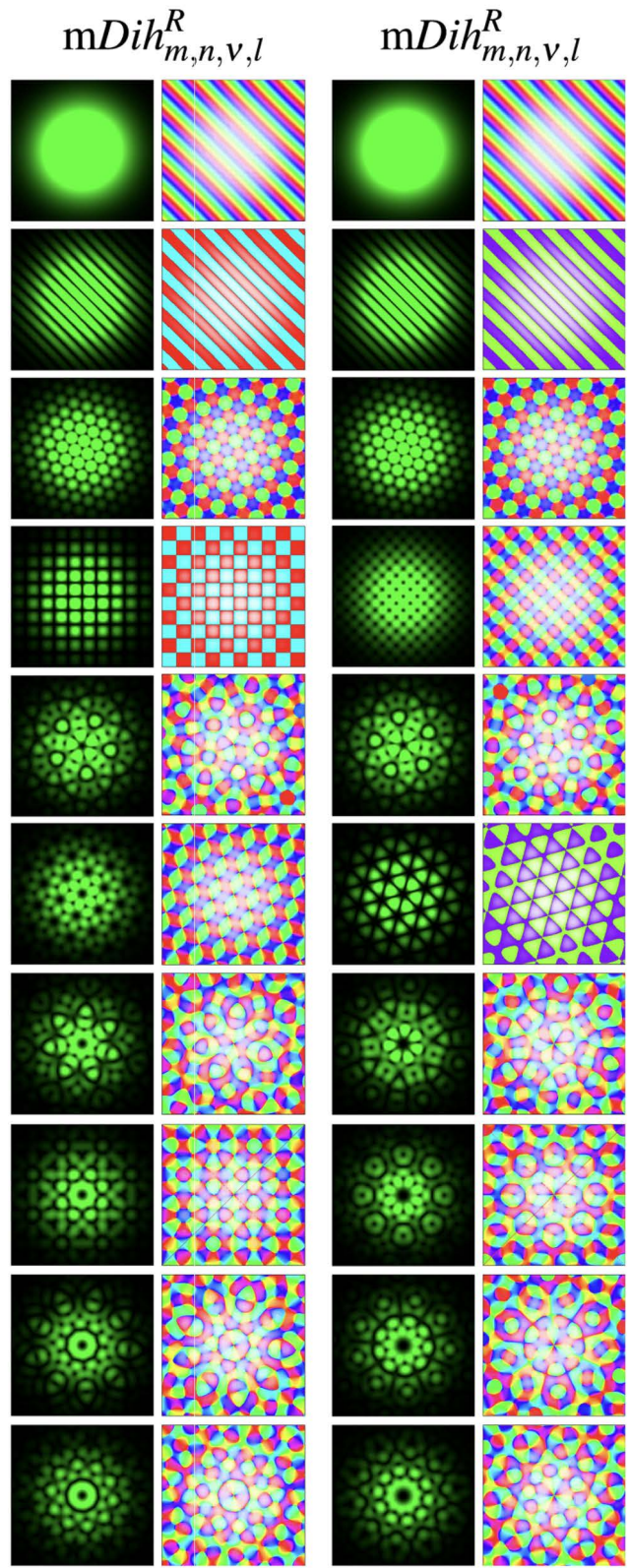

**Figure 13.** Transverse amplitude and phase distributions of the mDB2, $\mathrm{m}Dih^R_{m,n,\nu,l}$ for $l = 2$ (first two columns, resp.), and $l = 3$ (last two columns, resp.), for $m = n = 20$, at $z = 0$. The first row corresponds to $\nu = 1$, and so on, until the tenth row corresponds to $\nu = 10$, where $\nu$ determines the dihedral group $D_\nu$. The plot range of all the figures is $(-2.5, 2.5)$, in normalized units, for both axes.

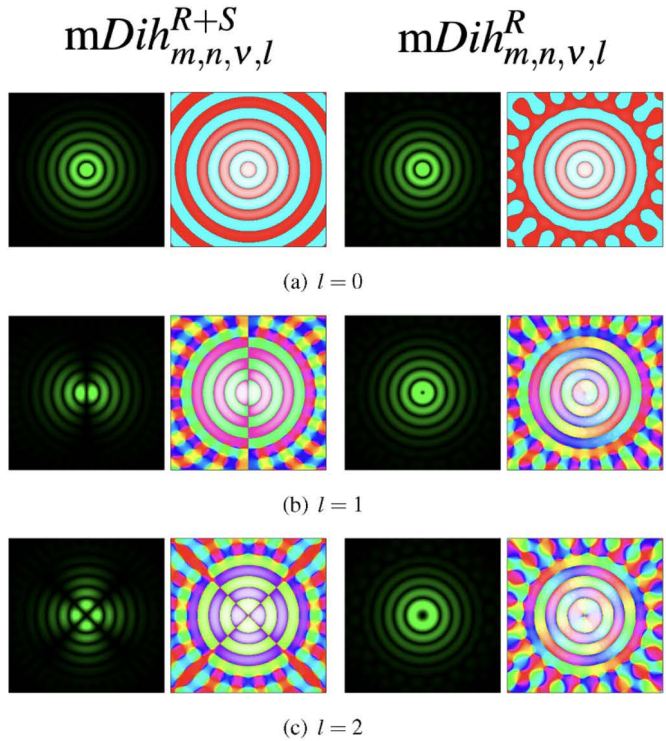

**Figure 14.** Transverse amplitude and phase distributions of the mDB1, $mDih_{m,n,\nu}^{R+S}$ (first two columns, resp.), and of the mDB2, $mDih_{m,n,\nu}^{R}$ (last two columns, resp.), for $\nu = 22$, $m = n = 20$, at $z = 0$, and (a) $l = 0$, (b) $l = 1$, and (c) $l = 2$. The plot range of all the figures is $(-2.5, 2.5)$, in normalized units, for both axes.

The general formulation presented in this work applies not only for scalar elegant beams but also for arbitrary vector input beams, and extends beyond the transformations of the dihedral group; as demonstrated in appendix C, it can be applied to any finite group of linear transformations. This formulation paves the way for new research in the creation and control of broader classes of structured light exhibiting a diverse kind of symmetries. Our approach can be useful in applications such as in the generation of optical lattices, and in optical micromanipulation, among others. Given the generality of our framework, it may impact other areas beyond optics and its quantum and mechanical analogs, as linear transformations are ubiquitous in physics.

## Data availability statement

No new data were created or analysed in this study.

## Acknowledgments

The author would like to acknowledge J A Ugalde-Ontiveros for valuable help in the generation of 3D visualization figures, as well as B Pérez-García for useful discussions. The author is also grateful for the financial support from the School of Engineering and Science, Tecnologico de Monterrey, to publish this article. Finally, the author would like to acknowledge I Jaimes-Loperena for illuminating communications.

## Appendix A

In this section, we express the Neumann-Hermite function in terms of the confluent hypergeometric function and show that the corresponding solutions to the paraxial equation are square-integrable.

The Neumann-Hermite function $NH_n(x)$ can be written as

$$NH_n(x) = \beta_n (-1)^n (a_n x)^{b_n} {}_1F_1\left(\frac{b_n - n}{2}, \frac{1 + 2b_n}{2}, x^2\right), \quad (1)$$

where $b_n = [1 + (-1)^n]/2$, and ${}_1F_1$ denotes the confluent hypergeometric function [67]. When $n$ is an even integer, $\beta_n = (-1)^{n/2} n!/(n/2)!$, whereas for $n$ odd, it is expressed as, $\beta_n = 2(-1)^{(n-1)/2} n!/([(n-1)/2]! a_n)$, and $c_n = e^{3\pi i(1-b_n)/2} \beta_n$.

To investigate the square-integrability of the corresponding elegant traveling-wave solutions to the paraxial equation, we use the asymptotics of the confluent hypergeometric equation for large values of the argument [67]. It is easy to prove that, for large values of $\xi$, the asymptotics of the Neumann-Hermite function can be expressed as

$$NH_n(\xi) \sim d_n \xi^{-(n+1+b_n)} e^{\xi^2}, \quad (2)$$

where $d_n$ is a constant. The Neumann-Hermite function was introduced to construct traveling-wave solutions to the paraxial equation in the context of sHGB [38]. In this case, the Gaussian envelope of sHGB is $e^{-\xi^2/2}$, which implies that the square-integrability depends on the term $NH_n(\xi)e^{-\xi^2/2}$, whose asymptotics reads

$$NH_n(\xi)e^{-\xi^2/2} \sim d_n \xi^{-(n+1+b_n)} e^{\xi^2/2}. \quad (3)$$

This is not only non-square-integrable but exponentially divergent, a fact that was interpreted as a mathematical artifact of the paraxial approximation [37, 38]. However, in the elegant case, the Gaussian envelope of eHGB is $e^{-\xi^2}$, which implies the square-integrability of the elegant traveling waves, since

$$NH_n(\xi)e^{-\xi^2} \sim d_n \xi^{-(n+1+b_n)}. \quad (4)$$

This is why we say that the elegant transverse structure is responsible for the square-integrability of the elegant traveling waves.

## Appendix B

In this section, we obtain the far-field expression of the elegant traveling waves. For this, we first note that the argument of the Hermite functions in equation (1) can be expanded for $z \gg 1$, as

$$\frac{x}{\sqrt{1+iz}} \sim \sqrt{1-iz}\,\xi, \quad (5)$$

where $\xi = x/z$. Therefore, we note that for large $z$, keeping $\xi$ constant, the modulus of the above term goes to infinity. Hence, the large argument asymptotic expressions of the Hermite functions $H_n$ and $NH_n$ can be obtained from the asymptotics of the confluent hypergeometric function [77]. After some algebra, we arrive at the following first-order asymptotic approximations,

$$H_n\left(\frac{x}{\sqrt{1+iz}}\right) \sim \left(2\sqrt{1-iz}\,\xi\right)^n, \quad (6)$$

$$NH_n\left(\frac{x}{\sqrt{1+iz}}\right) \sim -i\frac{\xi}{|\xi|}\left(2\sqrt{1-iz}\,\xi\right)^n, \quad (7)$$

for $n$ integer, where the approximation $n \gg 1$ was used employed to simplify the expressions further. Notice that the above asymptotic expressions are different only by the term $-i\xi/|\xi|$. With this, the asymptotic expressions of the elegant Hankel–Hermite functions can be written as

$$eHH_n^{(1,2)}\left(\frac{x}{\sqrt{1+iz}}\right) \sim \left(2\sqrt{1-iz}\right)^n \left(1 \pm \frac{\xi}{|\xi|}\right) e^{iz\xi^2} \xi^n e^{-\xi^2}. \quad (8)$$

Given that the function $\left(1 \pm \frac{\xi}{|\xi|}\right)$ equals zero for one side of the $\xi$-axis, so does the asymptotic expression in equation (8). This contrasts with the well-known fact that the 1D eHGB profile has two peaks at first order, one in each side of the $\xi$-axis (the far field expressions of $H_n\left(\frac{x}{\sqrt{1+iz}}\right) e^{-\left(\frac{x}{\sqrt{1+iz}}\right)^2}$, can be obtained by replacing the term $\left(1 \pm \frac{\xi}{|\xi|}\right)$ by 1 in equation (8)) [44].

Now, let us pay attention to the term $\xi^n e^{-\xi^2}$ in equation (8), and let us use the asymptotic expression derived by Porras et al [47], namely, $t^{2\nu+s}e^{-t^2} \sim \Gamma(\nu+1)\sqrt{\nu^{s-1}/(2\pi)}\,e^{-2(t-\sqrt{\nu})^2}$, for $\nu \gg s$. By taking $t = \xi$, $n = 2\nu + s$, and $s = 0$ for $n$ even, while $s = 1$ for $n$ odd, the first order asymptotic expression in equation (8) can be written, for any integer $n$, as

$$eHH_n^{(1,2)}\left(\frac{x}{\sqrt{1+iz}}\right) \sim (\pm 1)^n d_n (1-iz)^{n/2} e^{ix^2/z}$$
$$\times e^{-\left(\pm x - z\sqrt{n/2}\right)^2/(z/\sqrt{2})^2}, \quad (9)$$

where $d_n = (2^{n+1}/\sqrt{n\pi})\Gamma(n/2+1)$, and $\Gamma$ denotes the Gamma function [67]. Therefore, the far-field asymptotic expression of the elegant traveling beams takes the form shown in equation (4).

## Appendix C

In this section, we demonstrate that the sum of all linear transformations of a finite group acting on a function is invariant under any group transformation.

Let $G$ be a finite group of linear transformations, acting on a given function space, whose group operation is the composition of transformations. Then, $G$ can be written as $G = \{g_1, g_2, \cdots, g_n\}$, where $n$ is a fixed natural number. Let $a \in G$ be an arbitrary element of the group, and let $aG$ denote the set of all the compositions of $a$ with the elements of $G$, that is, $aG = \{ag_1, ag_2, \cdots, ag_n\}$. Naturally, $aG \subseteq G$. Suppose that $aG \neq G$. Consequently, there is $b \in G$ such that $b \notin aG$. In other words, $b \neq ag_k$, for each $k \in \{1, \cdots, n\}$. Of course, since $G$ is a group, the inverse of $a$ exists and belongs to $G$, i.e, $a^{-1} \in G$. Hence, $a^{-1}b \in G$, implying the existence of $k \in \{1, \cdots, n\}$ such that $a^{-1}b = g_k$, leading to a contradiction. Therefore $aG = G$. This implies that, for any $f$ in the function space,

$$a\sum_{k=1}^n g_k f = \sum_{k=1}^n ag_k f = \sum_{k=1}^n g_k f, \quad (10)$$

for each $a \in G$.

## ORCID iD

Alfonso Jaimes-Nájera  https://orcid.org/0000-0003-0911-2343